\documentclass[prd,twoside,twocolumn,longbibliography]{revtex4-2}
\usepackage{amsmath}
\usepackage{bbm}
\usepackage{amssymb}
\usepackage{amsfonts}

\renewcommand{\vec}{\textbf}
\newcommand{\ket}[1]{|#1\rangle}
\newcommand{\bra}[1]{\langle#1|}
\newcommand{\bracket}[2]{\langle#1|#2\rangle}

\newcommand{\fb}[1]{\mathsf{B}(#1)}
\newcommand{\fbt}[1]{\tilde{\mathsf{B}}(#1)}
\newcommand{\id}{{\mathbb{I}}}
\DeclareMathOperator{\tr}{Tr} 
\usepackage{subfig}
\usepackage{graphicx}
\usepackage{color}

\begin{document}

\title{Entanglement and Bell inequality violation in vector diboson systems produced 
in decays of spin-0 particles}

\author{Alexander Bernal}
\email{alexander.bernal@csic.es}
\affiliation{Instituto de F\'isica Te\'orica, IFT-UAM/CSIC, 
	Universidad Aut\'onoma de Madrid,\\
	Cantoblanco, 28049 Madrid, Spain}
\author{Pawe{\l}{} Caban}
\email{Pawel.Caban@uni.lodz.pl (corresponding author)}
\author{Jakub Rembieli\'nski}
\email{jaremb@uni.lodz.pl}
\affiliation{Department of Theoretical Physics,
	University of {\L}{\'o}d{\'z}\\
	Pomorska 149/153, PL-90-236 {\L}{\'o}d{\'z}, Poland}

\begin{abstract}
We discuss entanglement and the violation of the 
Collins--Gisin--Linden--Mas\-sar--Popescu (CGLMP) inequality in a system
of two vector bosons produced in the decay of a spin-0 particle. 
We assume the most general CPT conserving, Lorentz-
invariant coupling of the spin-0 particle with the daughter bosons.
We compute the most general two-boson density matrix obtained by averaging over
kinematical configurations with an appropriate probability distribution 
(which can be obtained when both bosons subsequently decay into fermion-antifermion).
We show that the two-boson state is entangled and violates the CGLMP inequality for
all values of the (anomalous) coupling constants and that in this case 
the state is entangled iff it can violate the CGLMP inequality.
As an exemplary process of this kind we use the decay $H\to ZZ$
with anomalous coupling.
\end{abstract}


\maketitle

\section{Introduction}

One of the most intriguing and fascinating aspects of quantum mechanics is quantum
nonlocality. Nonlocality is expressed by correlations of entangled quantum states 
of quantum objects. As was proven theoretically by John Stuart Bell \cite{cab_Bell1964}
nonlocality is an 
immanent aspect of quantum reality which was confirmed for elementary systems in series 
of correlation experiments in different scales of energy-momentum as well as different 
conditions for space-time localization
\cite{FC1972,ADR1982,cab_RKMSIMW2001,AWetal2009_josephson,PTetal2013,HBDetal2015,GVetal2015,SMetal2015}. 
Recently, the possibility of testing 
Bell-type inequality violation
under
extremal conditions arising in a decay of the Higgs boson has been put forward
\cite{Barr2021,BCR2022-Bell-vector-bosons,ASBCM2022_entanglement_HtoZZ,APBW2022_entanglement-weak-decays,Aguilar-Saavedra2022_entanglement_HtoWW,FFGM2023,BCR2023_HtoZZ-anomalous,FHM2024_SemileptonicHtoWW}.
Indeed, in the most interesting case of Higgs decay into two  gauge bosons ($ZZ$ or $WW$) 
we have possibility to measure quantum spin states of unstable  vector bosons. 
This is possible  by means of registration of their decay products (leptons) which are detected 
in polarization states determined by kinematics. The decay process of weak bosons takes place 
in the extremely short time $10^{-25}$ sec and at last one of decaying bosons is in 
a virtual state, out of the mass shell. Therefore, investigation of possible correlations 
of entangled vector bosons can provide a test of quantum 
mechanics under completely new conditions.

It should be stressed here that analysis of spin correlations in the relativistic
regime is hindered by problems with the definition of a proper spin observable for
a relativistic particle. Various operators have been proposed in the literature,
see, e.g., \cite{cab_Czachor1997_1,CRW_2009_strange_correl,CRW_2013_Dirac_spin_PRA,BAKG2014_Spin,FOZ2014_SpinOperatorQFT,GCB_2019PhysRevLett.123.090404} and references therein.
In our paper we used the so called Newton-Wigner spin operator corresponding to
a spin of a particle in its rest frame, we justified this choice in, e.g.,
\cite{BCR2022-Bell-vector-bosons}. 
We will also shortly comment this point in Conclusions. 

Einstein--Podolsky--Rosen (EPR) type correlation experiment
with relativistic vector bosons in the simplest scalar state 
was considered in \cite{CRW_2008_vector_bosons} for the first time.
The possibility of observing 
the violation of Clauser--Horn--Shimony--Holt (CHSH) and
Collins--Gisin--Linden--Mas\-sar--Popescu (CGLMP) ine\-qualities 
by $WW$ bosons arising in the actual decay $H \to WW$ for the first time was
analyzed in \cite{Barr2021} 
(notice, that  the first proper discussion of spin correlations in this decay 
was given in \cite{DD1977-HWW-correl}).
In \cite{BCR2022-Bell-vector-bosons} the possibility of violation of CHSH, Mermin and
CGLMP inequalities by a boson-antiboson system in the most general scalar state
(introduced in this context in \cite{Caban_2008_bosons_helicity})
was discussed.
In \cite{ASBCM2022_entanglement_HtoZZ} the entanglement
and violation of CGLMP inequality by a pair of $Z$ bosons 
arising in the decay $H\to ZZ$ was considered.
The authors of \cite{ASBCM2022_entanglement_HtoZZ} assumed 
the Standard Model interaction of $H$ with the daughter $Z$ bosons
and showed that $ZZ$ state produced in such a process is highly entanglement.
In our paper \cite{BCR2023_HtoZZ-anomalous} we also discussed
entanglement and the violation of CGLMP inequality by a $ZZ$ pair
arising in the decay $H\to ZZ$ but assuming anomalous (beyond
the Standard Model) structure of the vertex describing interaction of a Higgs
particle with two daughter bosons. 
It can be shown (compare e.g. \cite{ZK_2016,GMM_2007-HZZ})
that the amplitude corresponding to the 
most general Lorentz-invariant, CPT conserving coupling of the pseudoscalar/scalar 
particle $X$ with two vector bosons $V_1$, $V_2$ in the decay
\begin{equation}
X\to V_1 V_2
\label{XtoVV}
\end{equation}
depends on three parameters,
denoted in Eq.~(\ref{general-vertex}) by $v_1$, $v_2$ and $v_3$.
The Standard Model Higgs decay $H\to ZZ$ corresponds to $v_1=1$, $v_2=v_3=0$
while $v_3\not=0$ implies the possibility of CP violation and a pseudoscalar component
of $H$. In \cite{BCR2023_HtoZZ-anomalous} we considered anomalous coupling but 
limited ourselves to the case of a scalar Higgs ($v_1\not=0$, $v_2\not=0$, $v_3=0$).

In this paper we extend our analysis and discuss entanglement and 
CGLMP inequality violation in the state
of two vector bosons arising in the decay of a  pseudoscalar/scalar particle $X$
given in Eq.~(\ref{XtoVV}) assuming that both bosons decay into leptons.
As an example of such a process we use the decay of the Higgs particle
into a pair of $Z$ bosons: $H\to ZZ$.
That is, we assume that both anomalous couplings $v_2$ and $v_3$ are nonzero,
we also assume $v_1\not=0$ to have the possibility to use the actual decay
$H\to ZZ$ as an example.
Notice that even for $v_1=0$ we can apply the same methods, we shortly comment 
this point after Eq.~(\ref{normalization-basis}). 

Anomalous coupling parameters for the decay $H\to ZZ$
are constrained by measurements of Higgs
properties performed at the LHC
\cite{CMSCollab2019-H-anomalous-PhysRevD.99.112003},
they are also constrained from the theoretical point of view by perturbative unitarity.
We discuss this point in \ref{sec:AppB-bouns}.
In the present paper we do not limit values of anomalous couplings to these bounds
since we treat the Higgs decay as an exemplary process only, our considerations are more 
general.

It is worth noticing that the Higgs decay is not the only process which was proposed
as a test bed for exploring fundamental quantum properties like entanglement or
Bell-type inequality violation in high energy physics.
In fact, the first propositions of EPR-like experiments in particle physics were
put forward more then forty years ago
\cite{Tornqvist1981,Privitera1992}.
In recent years, the first system considered in this context was a system of top quarks
produced in colliders \cite{AN2021-top-quarks,FFP2021-PhysRevLett.127.161801,ASC2022_top-quarks,AMMM2022_tt-quarks,DGKN2023-Bell-entanglement-tt,SBMS2022_TopsLHC,AN2023-QDiscSteering-tt-LHC,SV2023_Entanglement-tt-SMEFT}.
Moreover, the only experimental observation of entanglement at such high energy scale
has been recently reported by the ATLAS collaboration at the LHC in a $tt$ system
\cite{ATLAS2023_tt-entanglement} and subsequently confirmed by the CMS collaboration
\cite{CMS2024_tt-entanglement}.

Other high energy processes have also been proposed in this context, including,
among other: various scattering processes
\cite{Morales2023-boson-scattering,SZ2023_Bell_2-2-scattering},
$B^0\bar{B}^0$ mesons \cite{TIetal2021-PhysRevD.104.056004},
$tW$ systems \cite{AguilarSaavedra2023_postdecay},
$WW$ pairs produced in electron-positron colliders
\cite{BCCZ2024_ObservablesForBellInWW}, 
$\tau\tau$ pairs \cite{EFMV2024_EntanglementBellTauTauBelle2}
and tripartite systems
\cite{SS2024_Skurai_Three-body-entanglement,AguilarSaavedra2024_Tripartite_HtoZZWW,Morales2024-tripartite}.

Furthermore, the possibility of detecting (or bounding) new physics effects 
with the help of quantum information techniques has been discussed
\cite{MSTV2024_QDetectionNewPhysTopLHC,AMMM2023-New-phys-diboson,FFGM2023_anomalous}.
For a recent review of the subject of quantum entanglement and Bell inequality
violation at colliders see \cite{BFFGM2024_QEntanglementBellViolationColliders}.

We use the standard units ($\hbar=c=1$, here $c$ denotes the velocity of light),
the Minkowski metric tensor 
$\eta=\mathrm{diag}(1,-1,-1,-1)$ and assume $\varepsilon_{0123}=1$.

\section{Decay of a  pseudoscalar/scalar particle into two vector bosons}

We consider here the decay (\ref{XtoVV}), where, in general, $V$ bosons can be 
off-shell. We will treat off-shell particles like on-shell ones with reduced invariant masses,
similarly as it was done in previous papers 
\cite{ASBCM2022_entanglement_HtoZZ,FFGM2023,GMM_2007-HZZ,ZK_2016}. 
Let us denote by $M$ the mass of the  pseudoscalar/scalar particle $X$
and by $k,m_1$ and $p,m_2$ the four-momenta
and invariant masses of the daughter particles.
The amplitude corresponding to the 
most general Lorentz-invariant, CPT conserving coupling of the (pseudo)sca\-lar 
particle with two vector bosons can be written as
(see e.g. \cite{ZK_2016,GMM_2007-HZZ} )
\begin{multline}
\mathcal{A}_{\lambda\sigma}(k,p) \propto
\Big[v_1 \eta_{\mu\nu} + v_2 (k+p)_\mu (k+p)_\nu \\
+v_3 \varepsilon_{\alpha\beta\mu\nu} (k+p)^\alpha (k-p)^\beta \Big]
e_{\lambda}^{\mu}(k) e_{\sigma}^{\nu}(p),
\label{general-vertex}
\end{multline}
where $\lambda,\sigma$ are spin projections of the final states, 
$v_1$, $v_2$, $v_3$ are three real coupling constants, and
$\varepsilon_{\alpha\beta\mu\nu}$ is a completely antisymmetric 
Levi-Civita tensor.
Moreover, amplitude $e_{\lambda}^\mu(q)$ for
the four-momentum $q=(q^0,\vec{q})$ with ${q^0}^2-{\vec{q}}^2=m^2$
reads \cite{CRW_2008_vector_bosons}
\begin{equation}
	e(q) = [e^{\mu}_{\sigma}(q)] = 
	\begin{pmatrix}
		\tfrac{\vec{q}^T}{m}\\
		\id + \tfrac{\vec{q}\otimes \vec{q}^T}{m(m+q^0)}
	\end{pmatrix}
	V^T, 
	\label{amplitude-e-explicit}
\end{equation}
with
\begin{equation}
	\label{matrix_V}
	V=\frac{1}{\sqrt{2}}
	\begin{pmatrix}
		-1 & i & 0 \\
		0 & 0 & \sqrt{2} \\
		1 & i & 0 \\
	\end{pmatrix}.
\end{equation}
These amplitudes fulfill standard transversality condition
\begin{equation}
	e^{\mu}_{\sigma}(q) q_\mu = 0.
	\label{transversality}
\end{equation}

We do not provide a Lagrangian of the interaction,
since the anomalous vertices are obtained from operators 
carrying a dimension larger than four.
Thus, it would be necessary to collect all dimension-six operators in the SMEFT 
(Standard Model Effective Field Theory) associated with this interaction
which is out of the scope of the present work.

For our exemplary decay $H\to ZZ$
the Standard Model interaction corresponds to $v_1=1$, $v_2=v_3=0$.
Therefore, since we want to use actual experimental values of masses 
of the Higgs particle and $Z$ bosons in our numerical examples,
from now on we will assume that $v_1\not=0$.
Moreover, we admit nonzero $v_2$ and $v_3$.
Note that experimental data regarding Higgs decay admit nonzero 
$v_2$ and $v_3$ but give strong bounds on their values
\cite{CMSCollab2019-H-anomalous-PhysRevD.99.112003},
we will discuss these bounds later on.

With these assumptions, the most general pure state of two vector bosons arising in 
the decay (\ref{XtoVV}) can be parametrized with the help of two parameters, 
$c$, $\tilde{c}$, as
\begin{multline}
\ket{\psi_{VV}(k,p)} = 
\Big[
\eta_{\mu\nu} + \tfrac{c}{(kp)}(k_\mu p_\nu + p_\mu k_\nu) \\
+ \tfrac{\tilde{c}}{(kp)} \varepsilon_{\alpha\beta\mu\nu}
(k+ p)^\alpha( k - p)^\beta
\Big]
e_{\lambda}^{\mu}(k) e_{\sigma}^{\nu}(p)
\ket{(k,\lambda);(p,\sigma)},
\label{pseudoscalar-state-general}
\end{multline}
where
\begin{equation}
c = (kp) \tfrac{v_2}{v_1},\quad
\tilde{c} = (kp) \tfrac{v_3}{v_1},
\label{c_tilde-c_def}
\end{equation}
and $\ket{(k,\lambda);(p,\sigma)}$ is the two-boson state, one boson with 
the four-momentum $k$ and spin projection along $z$ axis $\lambda$,
second one with the four-momentum $p$ and spin projection $\sigma$.
For $k\not=p$ states $\ket{(k,\lambda);(p,\sigma)}$ are orthonormal:
\begin{equation}
	\bracket{(k,\lambda);(p,\sigma)}{(k,\lambda^\prime);(p,\sigma^\prime)} 
	=  \delta_{\lambda\lambda^\prime} 
	\delta_{\sigma \sigma^\prime}.
	\label{normalization-basis}
\end{equation}

In this paper we consider the case $v_1\not=0$ but
even for $v_1=0$ we can apply the same methods. 
For instance, if $v_3=0$ as well then the state is always separable, 
while if $v_3\not=0$ one can define a single parameter, e.g. $(kp)v_2/v_3$, 
and perform very similar analysis.

The state (\ref{pseudoscalar-state-general}) is not normalized,
with the help of Eq.~(\ref{normalization-basis}) we find
\begin{multline}
\bracket{\psi_{VV}(k,p)}{\psi_{VV}(k,p)} = 
2 + \Big[
(1+c) \tfrac{(kp)}{m_1 m_2} - c \tfrac{m_1 m_2}{(kp)}
\Big]^2 \\
+ 8 \tilde{c}^2 \Big[
1 - \Big(
\tfrac{m_1 m_2}{(kp)}
\Big)^2
\Big].
\label{normalization}
\end{multline}

We will use center of mass (CM) frame for our further computations. 
The kinematics of the decay (\ref{XtoVV}) in the CM frame 
is briefly summarized in \ref{sec:app-kinematics}.
Using formulas from this Appendix we find that in the CM frame
normalization of the state $\ket{\psi_{VV}(k,p)}$
depends only on masses $M$, $m_1$, $m_2$ and the parameters $c$, $\tilde{c}$:
\begin{equation}
\bracket{\psi_{VV}(k,p)}{\psi_{VV}(k,p)}|_{CM}  
= 2 (1+\tilde{\kappa}^2) + \kappa^2,
\label{normalization_CM}
\end{equation}
where, in analogy with our previous paper \cite{BCR2023_HtoZZ-anomalous}, 
we have introduced the following notation
\begin{equation}
\kappa = \beta + c (\beta - 1/\beta), \quad
\tilde{\kappa} = 2 \tilde{c} \sqrt{1- 1/\beta^2}
\label{kappas}
\end{equation}
and $\beta$ is given in Eq.~(\ref{beta-def}).

The ranges of possible values of $\kappa$, $\tilde{\kappa}$ depend on the values 
of $c$, $\tilde{c}$, respectively:
\begin{align}
&\kappa \in (-\infty,1] && \textrm{for} && c\in(-\infty,-1),&&\\
&\kappa \in [0,1] && \textrm{for} && c=-1,&&\\
&\kappa \in [2\sqrt{-c(1+c)},\infty) && \textrm{for} && c\in(-1,-\tfrac{1}{2}),&&\\
&\kappa \in [1,\infty] && \textrm{for} && c\in[-\tfrac{1}{2},\infty),&&
\end{align}
and
\begin{align}
&\tilde{\kappa} \in (2\tilde{c},0] && \textrm{for} && \tilde{c}\in(-\infty,0),&&\\
&\tilde{\kappa} \in [0,2\tilde{c}] && \textrm{for} && \tilde{c}\in[0,\infty).&&
\end{align}
We have given the admissible ranges of $\kappa$, $\tilde{\kappa}$ for all
values of $c$, $\tilde{c}$.
However, for a real decay, like our exemplary process $H\to ZZ$, there exist 
further experimental and theoretical bounds on possible values of $c$, $\tilde{c}$;
for the mentioned process $H\to ZZ$ we discuss these bounds in \ref{sec:AppB-bouns}.

Next, without loss of generality we can assume that bosons arising in the decay 
(\ref{XtoVV}) move along $z$-axis, i.e. we can take 
$k^\mu=(\omega_1,\vec{k})$,
$p^\mu=(\omega_1,-\vec{k})$,
where $\vec{k}=(0,0,|\vec{k}|)$ and energies $\omega_1$, $\omega_2$
are given explicitly in (\ref{formula_3},\ref{formula_4}).
We also simplify the notation of basis two-boson states in this case
\begin{equation}
\ket{\lambda,\sigma} \equiv
\ket{(\omega_1,0,0,|\vec{k}|);(\omega_2,0,0,-|\vec{k}|)}.
\end{equation}
In this notation, with the help of Eqs.~(\ref{amplitude-e-explicit}, \ref{pseudoscalar-state-general}, \ref{normalization_CM}), the normalized
state of two bosons reads
\begin{multline}
\ket{\psi_{VV}^{\mathsf{norm}}(m_1, m_2, c, \tilde{c})} = 
\frac{1}{\sqrt{2 (1+\tilde{\kappa}^2) + \kappa^2}}
\Big[
(1-i\tilde{\kappa}) \ket{+,-}\\
 - \kappa \ket{0,0}
+(1+i\tilde{\kappa}) \ket{-,+}
\Big].
\end{multline}
It should be noted that when $\tilde{\kappa}=0$ the above state coincides with the state
discussed in our previous paper
\cite{BCR2023_HtoZZ-anomalous}.

Bosons arising in a single decay (\ref{XtoVV}) have definite masses $m_1$ and $m_2$;
thus two-boson state is pure and has the following form
\begin{equation}
\rho(m_1,m_2,c,\tilde{c}) 
= 
\ket{\psi_{VV}^{\mathsf{norm}}(m_1, m_2, c, \tilde{c})}
\bra{\psi_{VV}^{\mathsf{norm}}(m_1, m_2, c, \tilde{c})}.
\label{rho-pure}
\end{equation}
However, when one determines two-boson state from experimental data then
averaging over various kinematical configurations is necessary and the state 
becomes mixed
\begin{equation}
\rho_{VV}(c,\tilde{c}) = 
\int dm_1\, dm_2\, \mathcal{P}_{c,\tilde{c}}(m_1,m_2) \rho(m_1,m_2,c,\tilde{c}),
\label{rho_VV_mixed-def}
\end{equation}
where $\mathcal{P}_{c,\tilde{c}}(m_1,m_2)$ is a normalized probability distribution.
The explicit form of this probability distribution can be determined 
for different channels of the subsequent decay of daughter 
vector bosons $VV$ arising in (\ref{XtoVV}). 
In our exemplary case of the Higgs decay into Z bosons,
the decay chain 
$H\to Z Z^* \to (f^+ f^-)(f^+ f^-)$
constitutes one of the most promising channels to certify 
entanglement at colliders in a qutrit-qutrit system.
Thus we stick to this channel in our general analysis.
In the case when the daughter bosons decay into
massless fermions
\begin{equation}
	X \to VV \to f_{1}^{+} f_{1}^{-} f_{2}^{+} f_{2}^{-},
\end{equation}
we can use the results from \cite{ZK_2016,BCR2023_HtoZZ-anomalous} and
following exactly the same line of reasoning as in our previous paper
(compare Eqs.~(33--38) from \cite{BCR2023_HtoZZ-anomalous}) we find
\begin{equation}
\mathcal{P}_{c,\tilde{c}}(m_1,m_2) 
= N \frac{\lambda^{\frac{1}{2}}(M^2,m_1^2,m_2^2) m_1^3 m_2^3}{D(m_1) D(m_2)}
\big[2(1+\tilde{\kappa}^2)+\kappa^2\big],
\label{probability-distribution-final}
\end{equation} 
with
\begin{equation}
D(m)  = \big(m^2-m_V^2\big)^2 + (m_V \Gamma_V)^2,
\label{D-def}
\end{equation}
where $m_V, \Gamma_V$ denote the mass and decay width of the on-shell $V$ boson
and the normalization factor $N$ can be determined numerically for given values
$c$ and $\tilde{c}$.

Therefore, introducing the notation
\begin{align}
\fb{n} & = \int\limits_{S} \! dm_1 dm_2 \frac{\lambda^{1/2}(M^2,m_1^2,m_2^2)
m_1^3 m_2^3}{D(m_1) D(m_2)} \beta^n, 
\label{B-def}\\
\fbt{n} & = \int\limits_{S} \! dm_1 dm_2 \frac{\lambda^{1/2}(M^2,m_1^2,m_2^2)
	m_1^3 m_2^3}{D(m_1) D(m_2)} \beta^n (\beta^2 -1)^{1/2},
\label{Btilde-def}
\end{align}
for $n=-2,-1,0,1,2$, where $S=\{ (m_1,m_2): m_1\ge0, m_2\ge0, m_1+m_2 \le M \}$,
the state averaged over kinematical configurations (\ref{rho_VV_mixed-def}) can 
be written as
\begin{equation}
\rho_{VV}(c,\tilde{c}) = 
	\frac{1}{\mathsf{b}+2\mathsf{e}}
	\begin{pmatrix}
	0 & 0 & 0 & 0 & 0 & 0 & 0 & 0 & 0\\
	0 & 0 & 0 & 0 & 0 & 0 & 0 & 0 & 0\\
	0 & 0 & \fbox{$\mathsf{e}$} & 0 & \fbox{$\mathsf{f}$} & 0 & \fbox{$\mathsf{h}$} & 0 & 0\\
	0 & 0 & 0 & 0 & 0 & 0 & 0 & 0 & 0\\
	0 & 0 & \fbox{$\mathsf{f}^*$} & 0 & \fbox{$\mathsf{b}$} & 0 & \fbox{$\mathsf{f}$} & 0 & 0\\
	0 & 0 & 0& 0 & 0 & 0 & 0 & 0 & 0\\
	0 & 0 & \fbox{$\mathsf{h}^*$} & 0 & \fbox{$\mathsf{f}^*$} & 0 & \fbox{$\mathsf{e}$} & 0 & 0\\
	0 & 0 & 0 & 0 & 0 & 0 & 0 & 0 & 0\\
	0 & 0 & 0 & 0 & 0 & 0 & 0 & 0 & 0
\end{pmatrix},
\label{rhoVV-mixed-final}
\end{equation}
where for better visibility we have framed the non-zero matrix elements:
\begin{align}
\mathsf{b} & =  -2c(1+c)\fb{0}+ (1+c)^2\fb{2} + c^2 \fb{-2},
\label{b-def}\\
\mathsf{e}  & = (1+4\tilde{c}^2)\fb{0} - 4\tilde{c}^2 \fb{-2},
\label{e-def}\\
\mathsf{f} & = -(c+1)\fb{1} + c \fb{-1} 
+2i\tilde{c}\big[ (1+c) \fbt{0}
-c \fbt{-2} \big],
\label{f-def}\\
\mathsf{h} & = (1-4\tilde{c}^2) \fb{0} +4\tilde{c}^2 \fb{-2} -4i\tilde{c} \fbt{-1},
\label{h-def}
\end{align}
and star denotes complex conjugation.

When we want to obtain two-boson density matrix for
our exemplary decay $H\to ZZ$, we have to insert 
into Eqs.~(\ref{probability-distribution-final},\ref{D-def},\ref{B-def},\ref{Btilde-def})
the measured values for the Higgs mass, $Z$ mass and $Z$ decay width,
i.e., $M=M_H=125.25\; \mathsf{GeV}$, $m_V=m_Z=91.19\; \mathsf{GeV}$,
$\Gamma_V=\Gamma_Z=2.50\; \mathsf{GeV}$ \cite{Workman:2022-PDG}.
With these values, from (\ref{b-def},\ref{e-def},\ref{f-def},\ref{h-def})
we receive 
\begin{align}
\mathsf{b}_Z & =  9431.55 + 12883.6 c + 4983.07 c^2,
\label{bZ-explicit}\\
\mathsf{e}_Z  & = 2989.76 + 5834.84 \tilde{c}^2,
\label{eZ-explicit}\\
\mathsf{f}_Z & = -4819.07 - 2752.19 c +  7052.85 i \tilde{c} 
+ 4477.64 i c \tilde{c},
\label{fZ-explicit}\\
\mathsf{h}_Z & = 2989.76 - 8031.86 i \tilde{c} - 5834.84 \tilde{c}^2.
\label{hZ-explicit}
\end{align}

\section{Entanglement}
\label{sec:Entanglement}

To check whether the state (\ref{rhoVV-mixed-final}) is entangled 
and to estimate how much it is entangled we can use one of 
entanglement measures. In our previous paper \cite{BCR2023_HtoZZ-anomalous}
we used the logarithmic negativity
\cite{VW2002,Plenio_Logarithmic-negativity} which is a computable
entanglement measure and is defined as
\begin{equation}
	E_N(\rho_{AB}) = \log_3(||\rho^{T_B}||_1),
\end{equation}
where $T_B$ denotes partial transposition with respect to the subsystem $B$
and
$||A||_1=\tr(\sqrt{A^\dagger A})$ is the trace norm of a matrix $A$.
$||A||_1$ is equal to the sum of all the singular values of $A$;
when $A$ is Hermitian then it is equal to the sum of absolute values of all
eigenvalues of $A$.
$E_N(\rho)>0$ implies that the state $\rho$ is entangled.

It is worth noticing that the general structure (the number and positions of
non-zero entries) of the density matrix (\ref{rhoVV-mixed-final}) is the same
as the structure of the density matrix describing a $ZZ$ pair produced in the
decay of the Standard Model Higgs particle analyzed in
\cite{ASBCM2022_entanglement_HtoZZ}. In this paper it was shown that 
for a density matrix with such a structure the Peres--Horodecki criterion is
not only sufficient but also necessary for the state to be entangled.
And this implies that the state (\ref{rhoVV-mixed-final}) is entangled iff at 
least one off-diagonal matrix entry is non-zero.

\begin{figure}
\begin{center}
\includegraphics[width=0.9\columnwidth]{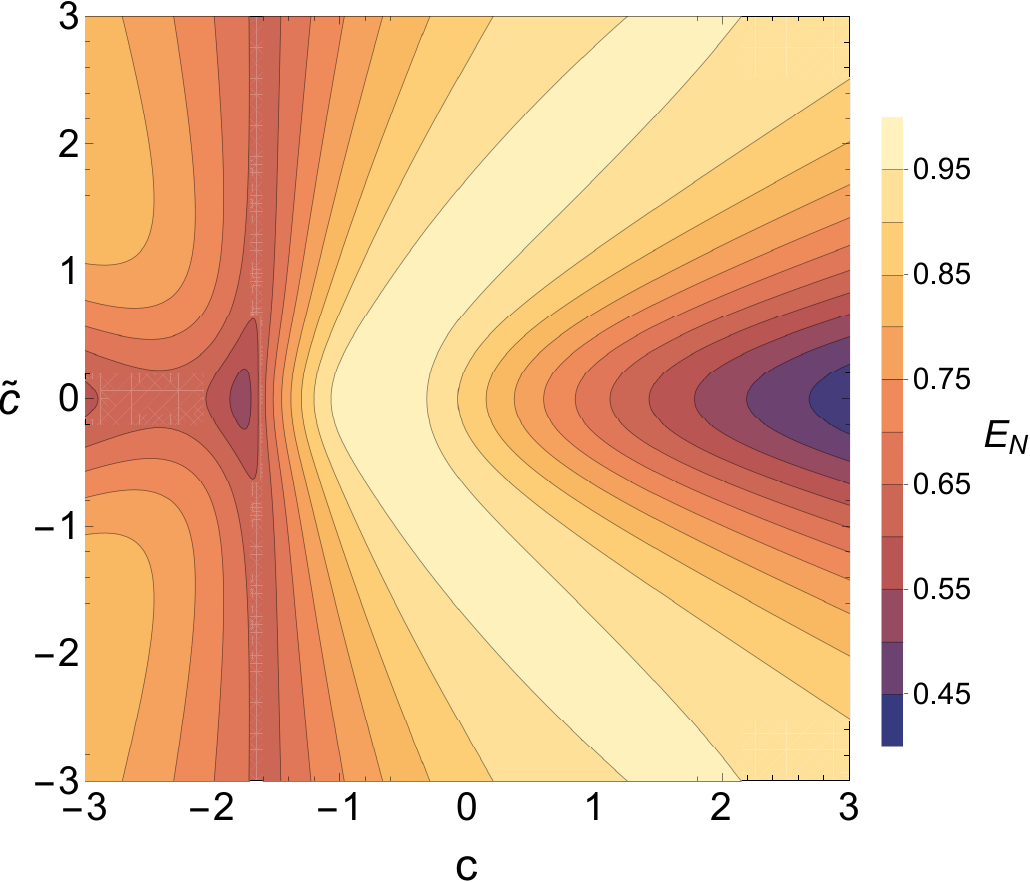}
\end{center}
\caption{In this figure we present  logarithmic negativity of the state
(\ref{rhoVV-mixed-final}), $E_N(\rho_{VV}(c,\tilde{c}))$ as a function of 
$c$, $\tilde{c}$. To obtain this plot 
we have inserted the measured values for the Higgs
mass, Z mass and Z decay width, i.e., we put 
$\mathsf{b}_Z$, $\mathsf{e}_Z$, $\mathsf{f}_Z$,
$\mathsf{h}_Z$ given in Eqs.~(\ref{bZ-explicit},\ref{eZ-explicit},\ref{fZ-explicit},\ref{hZ-explicit}).
}
\label{fig:negativity}	
\end{figure}

In Fig.~\ref{fig:negativity} we have plotted the logarithmic negativity of the state
(\ref{rhoVV-mixed-final}) for the decay $H\to ZZ$, i.e. 
with matrix elements $\mathsf{b}_Z$, $\mathsf{e}_Z$, $\mathsf{f}_Z$,
$\mathsf{h}_Z$ given in Eqs.~(\ref{bZ-explicit},\ref{eZ-explicit},\ref{fZ-explicit},\ref{hZ-explicit}).
In this case 
numerically obtained maximal value of the logarithmic negativity is equal to $0.99638$.
This value is attained for $c=-0.73719$, $\tilde{c}=0.00005$.
Moreover, $E_N>0$ for all values of $c$, $\tilde{c}$ and in the limit $c\to\infty$
the logarithmic negativity tends to zero.

\section{Violation of Bell inequalities}

Now, let us consider the violation of Bell inequalities in the state (\ref{rhoVV-mixed-final}).
The optimal Bell inequality for a two-qudit system was formulated in 
\cite{CGLMP_Bell_ineq_high_spin} and is known as the
Collins--Gisin--Linden--Massar--Popescu (CGLMP) inequality.
For two qubits ($d=2$) the CGLMP inequality reduces to the well known
Clauser--Horn--Shimony--Holt (CHSH) inequality \cite{cab_CHSH1969}.
Here we are interested in the CGLMP inequality for a two-qutrit system
(for spin-1 particle there are three possible outcomes of a spin projection
measurements). In this case the CGLMP inequality has the following form
\begin{equation}
\mathcal{I}_3 \le 2,
\label{CGLMP-def}
\end{equation}
where
\begin{multline}
	\mathcal{I}_3 = \big[
	P(A_1=B_1) + P(B_1=A_2+1)\\
	+ P(A_2=B_2) +P(B_2=A_1)
	\big]\\
	-\big[
	P(A_1=B_1-1) + P(B_1=A_2)\\
	+P(A_2=B_2-1) + P(B_2=A_1-1)
	\big]
\label{I3-def}
\end{multline}
and $A_1$, $A_2$ ($B_1$, $B_2$) are possible measurements that can be
performed by Alice (Bob). 
Each of these measurements can have three outcomes: 0,1,2.
Moreover, $P(A_i=B_j+k)$ denotes the probability 
that the outcomes $A_i$ and $B_j$ differ by $k$ modulo 3, 
i.e., $P(A_i=B_j+k) = \sum_{l=0}^{l=2} P(A_i=l,B_j=l+k \mod 3)$.
As usual, we assume that Alice can perform measurements on one of the bosons, 
Bob on the second one, i.e.,
we take Alice (Bob) observables as $A\otimes I$
($I\otimes B$).

To answer whether and how much a given quantum state $\rho$ violates 
the CGLMP inequality 
we have to find such observables $A_1$, $A_2$, $B_1$, $B_2$ for which the value
of ${\mathcal{I}}_3$ is maximal in the state $\rho$ (so called optimal observables).
But, in general, there does not exist a procedure of finding such optimal observables.  

The CGLMP inequality (\ref{CGLMP-def}) can be written as
\begin{equation}
	\tr\big( \rho \mathcal{O}_{\mathsf{Bell}}  \big) \le 2,
	\label{CGLMP-OBell}
\end{equation}
where $\mathcal{O}_{\mathsf{Bell}}$ is a certain operator depending on the
observables $A_1$, $A_2$, $B_1$, and $B_2$.

Each Hermitian $3\times3$ matrix $A$ can be represented with the help of
the $3\times3$ unitary matrix $U_A$, columns of $U_A$ are normalized 
eigenvectors of $A$ in a given basis.
Using this notation 
in \cite{ASBCM2022_entanglement_HtoZZ} it was shown that
\begin{multline}
\mathcal{O}_{\mathsf{Bell}}(U_{A_1},U_{A_2},U_{B_1},U_{B_2})=\\
-[U_{A_1}\otimes U_{B_1}] P_1 [I\otimes S^3] 
P_{1}^{\dagger} [U_{A_1}\otimes U_{B_1}]^\dagger\\
+ [U_{A_1}\otimes U_{B_2}] P_0 [I\otimes S^3] 
P_{0}^{\dagger} [U_{A_1}\otimes U_{B_2}]^\dagger\\
+ [U_{A_2}\otimes U_{B_1}] P_1 [I\otimes S^3] 
P_{1}^{\dagger} [U_{A_2}\otimes U_{B_1}]^\dagger\\
- [U_{A_2}\otimes U_{B_2}] P_1 [I\otimes S^3] 
P_{1}^{\dagger} [U_{A_2}\otimes U_{B_2}]^\dagger,
\label{OBell-by-U}
\end{multline}
where $S^3$ is the standard spin $z$ component matrix, $S^3=\mathrm{diag}(1,0,-1)$,
and $P_0$, $P_1$ are $3^2\times 3^2$ block-diagonal  permutation matrices:
\begin{equation}
P_n = 
\begin{pmatrix}
C^n & \mathcal{O} & \mathcal{O} \\
\mathcal{O} & C^{n+1} & \mathcal{O} \\
\mathcal{O} & \mathcal{O} & C^{n+2}
\end{pmatrix}, \quad n=0,1, 
\end{equation}
where $\mathcal{O}$ is the $3\times 3$ null matrix and
$C$ is the $3\times3$ cyclic permutation matrix
\begin{equation}
C = 
\begin{pmatrix}
0 & 0 & 1\\
1 & 0 & 0\\
0 & 1 & 0	
\end{pmatrix}.
\end{equation}
Each $U$ from (\ref{OBell-by-U}) can be taken as an element of $SU(3)$ group
which has 8 parameters.
Therefore, to perform the full optimization of $\mathcal{O}_{\mathsf{Bell}}$ for a given 
state one should optimize over the $32$ dimensional parameter space which is computationally
challenging.
Thus, usually, one applies a certain optimization procedure in order to find optimal observables.

In Appendix B of our previous paper \cite{BCR2023_HtoZZ-anomalous} we 
described in detail two such procedures we used
in the case $\tilde{c}=0$ (for the state of $Z$ bosons arising in the Higgs decay).
The first of these procedures, originally introduced 
in \cite{ASBCM2022_entanglement_HtoZZ}, worked very well for $c$ close to 0.
The second one, inspired by the proof of Theorem 2 
in \cite{PR1992-GenericQNonlocality}, allowed us to show that the CGLMP
inequality is violated for all $c$.
We will apply here this second procedure to show explicitly that CGLMP
inequality is violated for all $c$, $\tilde{c}$ for all states (\ref{rhoVV-mixed-final})
for which at least one off-diagonal element is non-zero (${\mathsf{f}}\not=0$
or ${\mathsf{h}}\not=0$).
To this end, let us notice that the density matrix (\ref{rhoVV-mixed-final})
can be written as
\begin{equation}
\rho_{VV}(c,\tilde{c}) = 
\begin{pmatrix}
	0 & 0 & 0 & 0 & 0 & 0 & 0 & 0 & 0\\
	0 & 0 & 0 & 0 & 0 & 0 & 0 & 0 & 0\\
	0 & 0 & \fbox{$a_{11}$} & 0 & \fbox{$a_{12}$} & 0 & \fbox{$a_{13}$} & 0 & 0\\
	0 & 0 & 0 & 0 & 0 & 0 & 0 & 0 & 0\\
	0 & 0 & \fbox{$a_{12}^*$} & 0 & \fbox{$a_{22}$} & 0 & \fbox{$a_{23}$} & 0 & 0\\
	0 & 0 & 0& 0 & 0 & 0 & 0 & 0 & 0\\
	0 & 0 & \fbox{$a_{13}^*$} & 0 & \fbox{$a_{23}^*$} & 0 & \fbox{$a_{33}$} & 0 & 0\\
	0 & 0 & 0 & 0 & 0 & 0 & 0 & 0 & 0\\
	0 & 0 & 0 & 0 & 0 & 0 & 0 & 0 & 0
\end{pmatrix},
\label{rhoVV-mixed-final-1}
\end{equation}
where $a_{11}+a_{22}+a_{33}=1$, $a_{11},a_{22}, a_{33}\in{\mathbb{R}}_{+}$.
Moreover, we define unitary matrices
\begin{equation}
U_V(t,\theta) = 
\begin{pmatrix}
\cos\tfrac{t}{2} & 0 & e^{i\theta}\sin\tfrac{t}{2}\\
0 & 1 & 0\\
-e^{-i\theta} \sin\tfrac{t}{2} & 0 & \cos\tfrac{t}{2}
\end{pmatrix}
\end{equation}
and
\begin{equation}
O_A = 
\begin{pmatrix}
0 & 0 & 1\\
0 & -1 & 0\\
1 & 0 & 0
\end{pmatrix}.
\label{U_V}
\end{equation}
Now, we calculate the mean value of the operator
\begin{equation}
(O_A \otimes I)
\mathcal{O}_{\mathsf{Bell}}(U_V(0,0),U_V(\tfrac{\pi}{2},0),U_V(t,\theta),U_V(-t,\theta))
(O_A \otimes I)
\label{Bell-optimized}
\end{equation}
in the state (\ref{rhoVV-mixed-final-1}). The result can be written as
\begin{equation}
{\mathcal{I}}_3 = 2 + \tfrac{3}{2}\big[
a(\cos{t} -1) - 2 |a_{13}| \cos(\alpha+\theta) \sin t
\big],
\label{I3-optimization}
\end{equation}
where we used the following notation:
\begin{equation}
a_{11}+a_{33}=a, \qquad a_{13} = |a_{13}| e^{i\alpha},
\end{equation}
and $a\ge0$ as a sum of diagonal elements of a positive semidefinite matrix.
The maximal value of (\ref{I3-optimization}) is attained for
\begin{multline}
\sin(\alpha+\theta) = 0, \qquad \cos(\alpha+\theta) = \pm 1,\\
\cos t = \tfrac{a}{\sqrt{a^2+4|a_{13}|^2}},\qquad
\sin t = \mp \tfrac{2|a_{13}|}{\sqrt{a^2+4|a_{13}|^2}},
\end{multline}
and is equal to
\begin{equation}
({\mathcal{I}}_3)_{\mathrm{max}} = 
2 + \tfrac{3}{2} \big[
\sqrt{a^2 + 4 |a_{13}|^2} -a
\big].
\end{equation}
Thus, we see that 
$({\mathcal{I}}_3)_{\mathrm{max}}>2$ for $|a_{13}|\not=0$.

To show that $({\mathcal{I}}_3)_{\mathrm{max}}>2$ for $a_{12}\not=0$
or $a_{23}\not=0$ 
we need to change the block structure of the unitary matrices $U_V$
(\ref{U_V}) putting a nontrivial $2\times2$ block in the upper-left corner
for $a_{12}\not=0$ or in the lower-right corner for $a_{23}\not=0$
(notice that in our case $a_{12}=a_{23}$). We do not present the detailed 
calculations here since they are similar to those given above.

Summarizing, we have shown that CGLMP inequality can be violated for all values of 
$c$, $\tilde{c}$ if at least one of the elements $a_{12}$, $a_{13}$, $a_{23}$ is non-zero.

It is very interesting that, as we have noticed in the Section \ref{sec:Entanglement},
the same condition holds for the state (\ref{rhoVV-mixed-final}) to be entangled.
In other words, we have shown that the state $\rho_{VV}(c,\tilde{c})$
(\ref{rhoVV-mixed-final}) violates the CGLMP inequality iff it is entangled.
It is a non-trivial observation since
for an arbitrary $3\times 3$ quantum state $\rho$ such a statement is true
only if $\rho$ is pure.

This also leads to strong phenomenological implications, since we have proven that 
in a pair of vector bosons one can indirectly test the violation of the Bell inequality 
by checking that the pair is entangled, which is a much easier experimental task. 
And this regardless of the interaction among the spin-0 particle and the gauge bosons
(provided that they are CPT conserving and Lorentz invariant), 
so no future new physics can change the statement of entanglement and Bell violation 
in this system.

\begin{figure}
\begin{center}
\includegraphics[width=0.9\columnwidth]{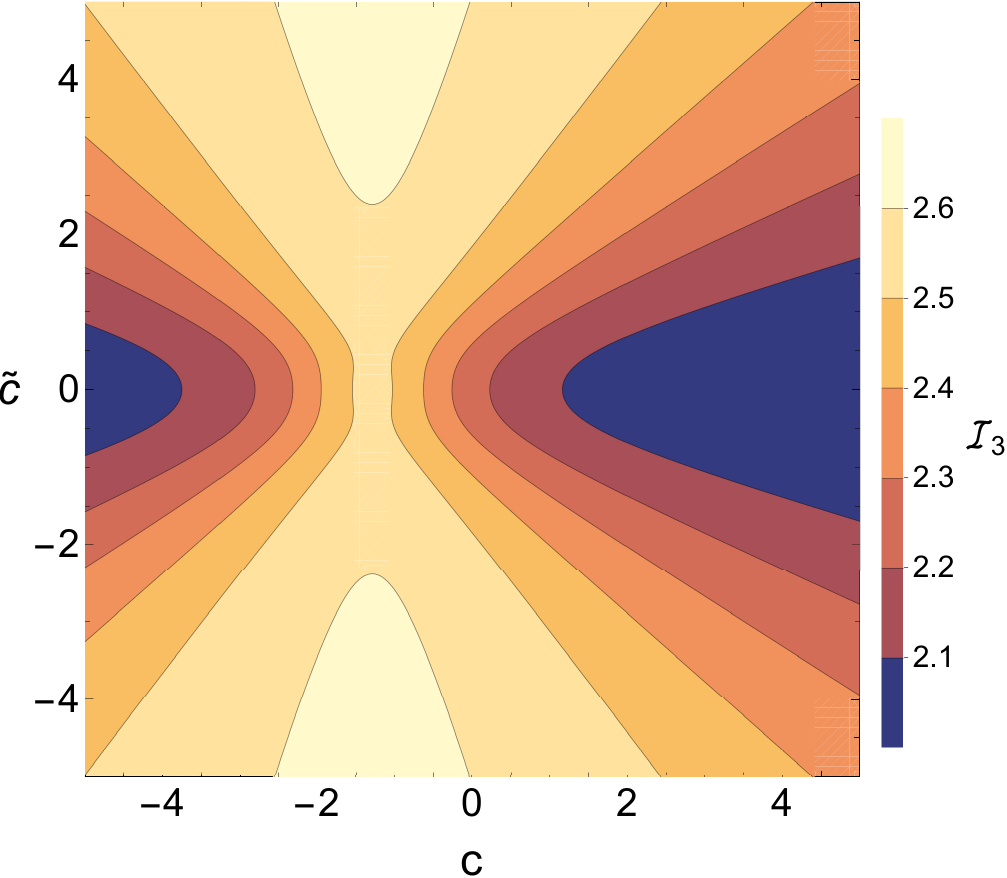}
\end{center}
\caption{In this figure we present the maximal value of $\mathcal{I}_3$
in the state (\ref{rhoVV-mixed-final}) as a function of $c$, $\tilde{c}$. 
We have inserted the measured values for the Higgs
mass, Z mass and Z decay width, i.e., we put $\mathsf{b}_Z$, $\mathsf{e}_Z$, 
$\mathsf{f}_Z$ and $\mathsf{h}_Z$ given in 
Eqs.~(\ref{bZ-explicit}--\ref{hZ-explicit}).
}
\label{fig:CGLMP}	
\end{figure}

\begin{figure}
	\begin{center}
		\includegraphics[width=0.9\columnwidth]{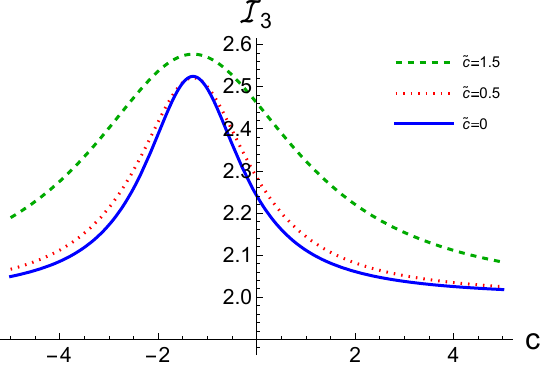}
	\end{center}
	\caption{In this figure we present the maximal value of $\mathcal{I}_3$
		in the state (\ref{rhoVV-mixed-final}) as a function of $c$, 
		for three chosen values of $\tilde{c}$. 
		We have inserted the measured values for the Higgs
		mass, Z mass and Z decay width, i.e., we put $\mathsf{b}_Z$, $\mathsf{e}_Z$, 
		$\mathsf{f}_Z$ and $\mathsf{h}_Z$ given in 
		Eqs.~(\ref{bZ-explicit}--\ref{hZ-explicit}).
		Notice that the choice $\tilde{c}=0$ corresponds to 
		the Standard Model value, $\tilde{c}=0.5$ to the experimental bound
		(\ref{ctildeHZZmax}). The value $\tilde{c}=1.5$ is presented for 
		comparison.	}
	\label{fig:CGLMP-cut}	
\end{figure}

In our exemplary decay $H\to ZZ$ from Eqs.~(\ref{bZ-explicit}--\ref{hZ-explicit})
we see that $a_{13}\not=0$. Therefore, in this decay the CGLMP inequality can be
violated for all values of $c$, $\tilde{c}$. 
In Fig.~\ref{fig:CGLMP} we have plotted the value of 
$({\mathcal{I}}_3)_{\mathrm{max}}$ obtained with the help of the 
above optimization procedure for that decay.
In Fig.~\ref{fig:CGLMP-cut} we have plotted 
$({\mathcal{I}}_3)_{\mathrm{max}}$ as a function of $c$ for three chosen values
of $\tilde{c}=0,0.5,1.5$. The value $\tilde{c}=0$ corresponds to the scalar Higgs,
$\tilde{c}=0.5$ corresponds to the boundary value given in Eq.~(\ref{ctildeHZZmax}) while
$\tilde{c}=1.5$ is given for comparison.

Comparing plots presented in Figs.~\ref{fig:negativity} and \ref{fig:CGLMP}
we see that state with the highest entanglement do not correspond to the 
state with the highest violation of the CGLMP inequality.
This observation is consistent with the general property of CGLMP inequality
\cite{ADGL2002_PhysRevA.65.052325}.

Experimentally, the state $\rho_{ZZ}(c,\tilde{c})$ is reconstructed in collider 
experiments via quantum tomography methods
\cite{APBW2022_entanglement-weak-decays,Bernal2023-tomography}. 
In such a case the presence of errors and background in the process
$H\to ZZ \to f_1^+ f_1^- f_2^+ f_2^-$
modifies the state (\ref{rhoVV-mixed-final}).
To estimate how this modification influences the violation of the CGLMP inequality
we consider the resistance of this violation with respect to the white noise.
It is worth noticing that the addition of white noise can also model,
in the first approximation, decoherence of the final state.
The noise resistance we define as a minimal value of $\lambda$, 
$\lambda_{\mathsf{min}}$, for which the state
\begin{equation}
\lambda \rho_{ZZ}(c,\tilde{c}) + (1-\lambda) \tfrac{1}{9} I_9,\qquad\lambda \in (0, 1]
\label{rho_ZZ-with-noise}
\end{equation}
violates the CGLMP inequality. 
Inserting the state (\ref{rho_ZZ-with-noise}) into the
CGLMP inequality (\ref{CGLMP-OBell}) we obtain
\begin{equation}
\lambda_{\mathsf{min}} = 
\frac{2}{\max\{\tr(\rho_{ZZ}(c,\tilde{c})\mathcal{O}_{\mathsf{Bell}}) \}}.
\label{lambda_min}
\end{equation}

\begin{figure}
\begin{center}
\includegraphics[width=0.9\columnwidth]{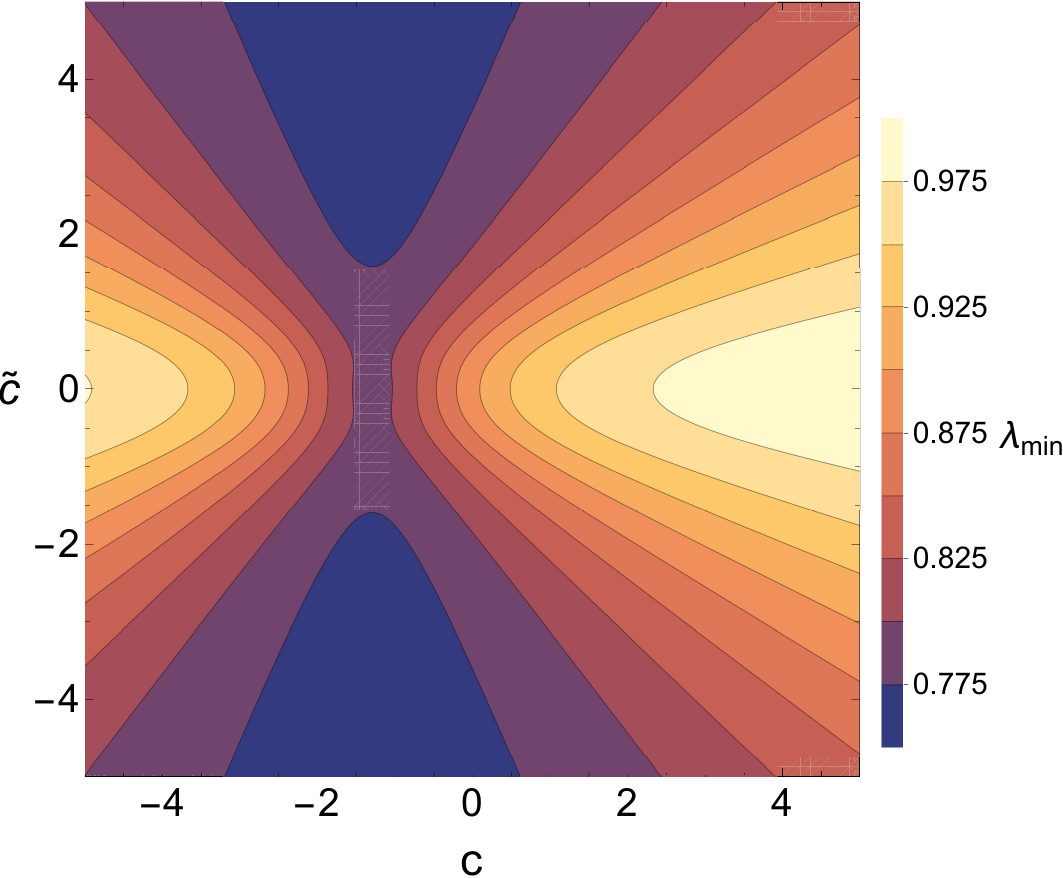}
\end{center}
\caption{In this figure we present  
$\lambda_{\mathsf{min}}$
(\ref{lambda_min}), as a function of $c$ and $\tilde{c}$.
We have inserted the measured values for the Higgs
mass, Z mass and Z decay width, i.e., we put $\mathsf{b}_Z$, $\mathsf{e}_Z$, 
$\mathsf{f}_Z$ and $\mathsf{h}_Z$ given in 
Eqs.~(\ref{bZ-explicit}--\ref{hZ-explicit}).}
\label{fig:noise}	
\end{figure}

We have plotted this value in Fig.~\ref{fig:noise} assuming that 
$\max\{\tr(\rho_{ZZ}(c,\tilde{c})\mathcal{O}_{\mathsf{Bell}}) \}$
is calculated with the help of the optimization procedure below Eq.~(\ref{Bell-optimized}).
From this plot we can see that for values $c$, $\tilde{c}$ close to 0 
we can tolerate up to almost a 20\% of noise and still attain a violation of 
the CGLMP inequality and hence an entangled state.

As we mentioned before, there exist other optimization procedures then the one 
used in this paper.  For example, one can focus on specific region
of the $(c,\tilde{c})$ parameter space (e.g. close to $(0,0)$).
Using such procedures one can attain larger violation 
of the CGLMP inequality in the considered region, which in consequence leads 
to higher noise tolerance.

Notice that values of $c$, $\tilde{c}$ close to 0 are expected for the decay
$H\to ZZ$ due to experimental bounds on anomalous couplings
for the $HZZ$ vertex \cite{RRS2021-anomalous-gauge-H}.
We discuss these bounds in more details in \ref{sec:AppB-bouns}.

\section{Conclusions}

We have discussed the CGLMP inequality violation and entanglement in a system 
of two vector bosons $V_1 V_2$ produced in the decay of a pseudoscalar/scalar 
particle $X$. 
As an example of such a process we use the decay of the Higgs particle into two $Z$
bosons.
We have assumed the most general CPT conserving, Lorentz-invariant
coupling of the particle $X$ with the daughter bosons $V_1 V_2$ 
(compare Eq.~(\ref{general-vertex})).
The amplitude of such a coupling depends on three parameters
$v_1$, $v_2$, $v_3$.
In the case of $H\to ZZ$, the Standard Model interaction corresponds to
$v_1=1$, $v_2=v_3=0$. On the other hand $v_3\not=0$ implies the possibility 
of CP violation and a pseudoscalar component of $H$.
Thus, we have assumed that $v_1\not=0$.
In such a case, the state of produced bosons,
beyond four-momenta and spins, can be
characterized by two parameters $c$, $\tilde{c}$ which, up to normalization are equal
to $v_2/v_1$ and $v_3/v_1$, respectively (cf. Eq.~(\ref{c_tilde-c_def})).
Next, in the center-of-mass frame, we have determined 
the most general pure state of $V_1 V_2$ boson pair for a particular event $X\to V_1 V_2$
and the $V_1 V_2$ density matrix $\rho_{VV}(c,\tilde{c})$ obtained by averaging 
over kinematical 
configurations with an appropriate probability distribution (which can be obtained when 
both bosons subsequently decay into leptons).
Finally, we have shown that this matrix is entangled and violates the CGLMP
inequality for all values of  $c$ and $\tilde{c}$ if at least one of off-diagonal
elements of the density matrix is non-zero.

In Introduction we noticed that different spin operators have been proposed
in the literature. In this context it is interesting to mention the paper 
\cite{FOZ2014_SpinOperatorQFT} in which spin correlations of an 
electron-positron pair arising in a decay of a pseudo-scalar particle were discussed.
The authors of \cite{FOZ2014_SpinOperatorQFT} use different
spin operator from the operator used in the present work. As a result 
they obtain that the considered $e^+ e^-$ pair becames unentangled
in the high energy limit.

It is also worth noticing here that spin correlation functions in colliders are
determined from cross-sections (or angular distribution of momenta).
This allows for construction of peculiar local hidden variable models 
duplicating experimental statistics 
(\cite{ADD1992_Bell-colliders,LSY2024_CanBeBellViolatedAtColliders}).
One can hope that this loophole will be closed with future technical developments
as it was the case in the standard low-energy Bell tests
\cite{cab_Genovese2005,HBDetal2015,Giustina_etal2013_Bell_test}.
The problem of loophole-free tests of Bell nonlocality at colliders
 has been recently addressed in
\cite{FFM2025-TestingBellColliders}. 
The authors of this paper take a different point of view.
They point out that by performing the quantum tomography 
of the system we are able to reconstruct the full two-partite density matrix. 
Then, any realization of the observables $A_1$, $A_2$, $B_1$, $B_2$
over the density matrix can be theoretically performed.

In this context
it is very interesting that, 
as we have shown, the state $\rho_{VV}(c,\tilde{c})$
(\ref{rhoVV-mixed-final}) violates the CGLMP inequality iff it is entangled.
It is a non-trivial observation since
for an arbitrary $3\times 3$ quantum state $\rho$ such a statement is true
only if $\rho$ is pure.
This also leads to strong phenomenological implications, since we have proven that 
in a pair of vector bosons one can indirectly test the Bell-type inequality violation
by checking that the pair is entangled.
And this seems to be a much easier experimental task
(compare the recent paper \cite{ATLAS2023_tt-entanglement}). 
Moreover, this observation holds regardless of the interaction among the spin-0 particle 
and the gauge bosons (provided that they are CPT conserving and Lorentz invariant, 
a very sensible requirement), 
so no future new physics can change the statement of entanglement and Bell violation 
in this system.

It is also important to stress that this relation between entanglement and the violation
of the CGLMP inequality is solely based on the texture of the matrix 
(which is a consequence of the symmetries involved in the decay) and not 
on the experimental way of getting the density matrix itself. 

In the paper we have considered the case $v_1\not=0$ in order to compare the results 
with the actual decay $H\to ZZ$. 
However, even for $v_1=0$ we can use the same methods and obtain similar results. 
For instance, if $v_3=0$ as well then the state is always separable, 
while if $v_3\not=0$ one can define a single parameter, e.g. $(kp)v_2/v_3$, 
and perform an analogous analysis.

Finally, In our paper as an exemplary process we considered the decay $H\to ZZ$.
However, some scalar or pseudo-scalar mesons could also decay in a similar way. 
For scalar mesons $\tilde{c}=0$ but one could have $c\not=0$ while
for pseudo-scalar ones $\tilde{c}\not=0$.

\begin{acknowledgements}
	P.C. and J.R. are supported by the University of Lodz.
	A.B. is grateful to J.A. Casas  and J. M. Moreno for very useful discussions. 
	A.B. acknowledges the support of the Spanish Agencia Estatal de Investigacion 
	through the grants ``IFT Centro de Excelencia Severo Ochoa CEX2020-001007-S''
	and PID2019-110058GB-C22 funded by\\
	MCIN/AEI/10.13039/501100011033 and by ERDF. 
	The work of A.B. is supported through the FPI grant PRE2020-095867 funded 
	by MCIN/AEI/10.13039/501100011033. 
\end{acknowledgements}

\appendix
\section{Kinematics of the decay $X\to VV$ in the center of mass frame}
\label{sec:app-kinematics}

We assume that the  pseudoscalar/scalar particle $X$ in the CM frame has
the four-momentum $(M,\vec{0})$ and decays into two, possibly off-shell, $V$
bosons with four-momenta
$k^\mu=(\omega_1,\vec{k})$, $\omega_1^2 - \vec{k}^2 = m_1^2$ and
$p^\mu=(\omega_2,-\vec{k})$, $\omega_2^2 - \vec{k}^2 = m_2^2$.
From energy conservation we get
\begin{equation}
M = \omega_1 +\omega_2,
\label{energy_conservation}
\end{equation}
and consequently, in the CM frame we have 
\begin{align}
\vec{k}^2 & = \frac{1}{4M^2} \lambda(M^2,m_1^2,m_2^2), 
\label{formula_1}\\
kp & = \frac{1}{2} \Big[M^2 - m_1^2 - m_2^2\Big], 
\label{formula_2}\\
\omega_1 & = \frac{1}{2M} \Big[M^2 + (m_1^2 - m_2^2)\Big], 
\label{formula_3}\\
\omega_2 & = \frac{1}{2M} \Big[M^2 - (m_1^2 - m_2^2)\Big],
\label{formula_4}
\end{align}
where, following e.g. \cite{ZK_2016} we have defined the following function
\begin{equation}
\lambda(x,y,z) = x^2+y^2+z^2 - 2xy - 2xz - 2yz.
\label{lambda-def}
\end{equation}
It is also convenient to introduce the following notation
\begin{equation}
\beta = \left. \tfrac{(kp)}{m_1 m_2}\right|_{CM}
=\frac{M^2 - (m_1^2 + m_2^2)}{2 m_1 m_2}.
\label{beta-def}
\end{equation}

\section{Experimental and theoretical bounds on $c$, $\tilde{c}$ for
the process $H\to ZZ$}
\label{sec:AppB-bouns}

For our exemplary decay $H\to ZZ$ there exist experimental and theoretical bounds
on anomalous couplings $v_2$ and $v_3$ in the vertex (\ref{general-vertex}).
These bounds imply bounds on our parameters $c$, $\tilde{c}$. 
We discuss them briefly here.

Strong experimental bounds come from the 
measurements of Higgs boson particles performed at the LHC by the CMS
Collaboration \cite{CMSCollab2019-H-anomalous-PhysRevD.99.112003}.
Comparing (2) in \cite{CMSCollab2019-H-anomalous-PhysRevD.99.112003}
and our Eq.~(\ref{general-vertex}) we obtain the following relation between
the parameterizations used in our paper and in the CMS Collaboration paper:
\begin{align}
	v_1 & \propto a_{1}^{ZZ} m_{Z}^{2} + 2 a_{2}^{ZZ} (kp),\\
	v_2 & \propto -2a_{2}^{ZZ},\\
	v_3 & \propto a_{3}^{ZZ}
\end{align}
(with the same proportionality constant).
In our previous paper \cite{BCR2023_HtoZZ-anomalous} we have discussed the bounds 
on $c$, we obtained
\begin{equation}
|c| < c_{\mathsf{HZZ}}^{\mathsf{max}} = 0.23.
\label{cHZZmax}
\end{equation}
Using similar argumentation for $\tilde{c}$ we obtain
\begin{equation}
|\tilde{c}| < \tilde{c}_{\mathsf{HZZ}}^{\mathsf{max}} = 0.5.
\label{ctildeHZZmax}
\end{equation}

From the theoretical point of view, perturbative unitarity 
(see, e.g., \cite{Logan2022_Perturbative-unitarity-Higgs} for a recent review)
can also constrain anomalous couplings. 
Using data from \cite{DDI2016_Perturbative-unitarity-anomalous},
we have shown in \cite{BCR2023_HtoZZ-anomalous} that
the requirement of perturbative unitarity does
not limit accessible values of $c$ in the process $H\to ZZ$.
We are not aware of any papers where perturbative unitarity is used to bound $v_3$.


%

\end{document}